# A gapless micro-dielectric-barrier-discharge ion source for analytical applications

Stephen L. Coy [a,b], Evgeny V. Krylov [a], Gary A. Eiceman [d], Isik Kanik [c]

([a]Imaginative Technologies LLC, Monrovia, CA;  [b]Dept. of Chemistry and Chemical Biology, Northeastern University, Boston, MA;  [c]Jet Propulsion Laboratory, 4800 Oak Grove Drive, Pasadena, California 91109; [d]Department of Chemistry and Biochemistry, New Mexico State University, Las Cruces, New Mexico 88003 )

## Abstract

Use of dielectric barrier discharge (DBD) as an ion source for sensitive chemical analysis by ion mobility, differential ion mobility, mass spectrometry or other techniques is uncommon because barrier discharges generate excess noise due to spatial and temporal instability.  This design uses contacted, crossed glass-coated micro-wires to focus the field into a gradually vanishing gap, suppressing spatial and temporal variability, reducing pressure, temperature, and humidity effects, stabilizing discharge initiation and limiting chemical fragmentation. Positive-ion-mode proton-transfer, chemical fragmentation from a micro-discharge, and $NO^+$ adducts combine to allow a broad range of applications. We analyze noise properties of the ion source and report chemical responsivity for a wide range of volatile organic compounds. Source noise spectral density is compared for three systems: the contacted coated wires source, a gapped dielectric barrier discharge source, and a 5 mCi Ni-63 radioactive source. The crossed-wires source shows noise properties approaching those of the white-noise Ni-63 source, while gapped discharge exhibits 1/f noise from area-discharge random path and intensity variations.  For chemical sensitivity testing, dilute samples are delivered by vapor flow injection or by gas chromatography, and then detected by differential ion mobility spectrometry (DMS / FAIMS) and, in a few cases by mass spectrometry. The compounds tested in positive ion mode include ketones and alcohols, simple aromatics (benzene, toluene, xylenes, ethyl and propyl benzene), chlorinated and nitrated solvents and aliphatics (hexanes and n-octane), with sensitivities from ppt to ppb levels.  The wires source for trace vapor detection is stable under a range of environmental conditions from very low (ppb) humidity levels upward, has wider chemical coverage than the radioactive nickel source with nearly equivalent noise properties, and provides an adjustable in intensity that can greatly exceed a radioactive source.

# Contents

# Introduction

Dielectric barrier discharges make use of a high frequency field applied through a protective and current-limiting dielectric barrier to develop a plasma discharge in the gap between dielectrics. Dielectric barrier discharges have been studied extensively, but not used successfully for analytical applications. A brief search of the scientific and technical literature on the term can return thousands of references, Very few of these are for analytical applications, with most discussing destruction of chemical waste, with a few more esoteric applications, such as ion beams for space propulsion, making an appearance.

A primary limitation of dielectric barrier discharge for analytical applications resides in the inherent fluctuations in path and current due to the repeated start and stop of the discharge current in multiple filaments. These fluctuations lead to noise in the intensity of the ion stream. These fluctuations are not captured in a one-dimensional model of breakdown, but result from the inherently two- or three-dimensional nature of discharge initiation. The initiation phase depends on surface charges and roughness, and other factors including local gas composition, pressure and temperature, gas-phase chemistry and ion-neutral scattering and reaction.

In this paper, we describe a configuration that focuses the electric field into a decreasing, and finally gapless, region of contact created at the crossing of glass-coated micro-wires. The noise characteristics of this glass-wires source are compared with those of a gapped dielectric barrier discharge between planar ceramics, and those of a Ni-63 radioactive source such as those commonly used in atmospheric pressure ionization for ion mobility spectrometry (IMS) and differential mobility spectrometry (DMS / FAIMS). The micro-wire source is found to have noise characteristics approaching the Ni-63 source, which shows frequency-independent white noise, while the gapped discharge exhibits 1/f or shot / flicker noise behavior that increases at lower frequency and interferes with rapid data acquisition. The wires source is similar to the device described by Krylov [1], which was specialized for argon detection. An earlier publication by Wentworth et al. [2] used a gapped discharge in helium and so created metastable helium for Penning ionization, as in the commercial DART ion source [3, 4]. Our work is the first detailed examination of noise properties and broad chemical testing for a fast, sensitive, low noise, adjustable intensity, dielectric barrier discharge source suitable for analytical applications.

For testing of chemical response, we couple the source to a differential mobility spectrometer (DMS) detector. The DMS detector detects only ions of selected ion-mobility field dependence, allowing separate selection of distinct ion species, or detection of all ions when operated in "transparent mode" with no fields applied. DMS, which is also known as FAIMS (field-asymmetric IMS) filters ion based on the averaged difference between high and low field ion mobility coefficients. Background information on the differential mobility technique is available in references such as and the monographs by Shvartsburg [5], and Eiceman, Hill, and Karpas[6] , or in other articles by the several groups developing advanced DMS applications [7-12].

## Experimental Section

This experimental section includes a description of the design and electrical properties of the crossed-wires and gapped discharge ion sources with DMS detection, and of the methods used for acquisition of noise power spectral data and chemical testing.

### Design of Discharge Sources

The use of dielectric barrier discharge as an ion source for sensitive chemical analysis methods such as ion mobility, differential ion mobility, or mass spectrometry, faces several challenges. For analytical applications, low noise, high speed, and good ionization efficiency are needed, along with broad chemical coverage, low fragmentation of the analyte, and low operating power. Also, due to the regulatory requirements faced in the use of radioactive sources such as Ni-63, alternative atmospheric pressure ion sources can be preferred in many applications. In a dielectric barrier discharge, current and voltage are reversed during each cycle the driving waveform. Since discharge starting currents can require a high overvoltage, these starting transients can deposit large amounts of energy, creating excessive fragmentation and reactive chemical species which complicate the spectrum and which may be corrosive to the instrument itself.

The dynamical stochastic behaviors of a dielectric barrier discharge are both temporal and spatial. In the time domain, initiation of a gapped dielectric barrier discharge has been studied by a number of authors. The results of needle to dielectric-coated plane by Emelyanov and Shemet [13] show that the current in the needle to barrier configuration jumps in a few nanoseconds to a peak and decays rapidly with a time constant of about 20 ns., and that the peak current, which is low when the needle is in contact with the barrier, increases rapidly when the gap increases to 100 um and then decreases again at still larger gaps. Similar time-dependence results are also shown by Shao et al. [14]. The potentially destructive impact energy of charges on the opposing electrode is only limited by gas-phase ion-neutral scattering, a process which has a mean free path of about 5 um at one atmosphere (Iza et al. [15] ). In that work, microwave frequency excitation of a 25 um gap allowed collisional cooling of the discharge, and resulted in a long device lifetime and a nearly neutral plasma. This device was operated in continuous mode generating a stable point micro-plasma. Attempting to operate it in pulsed mode required higher power and lead to device damage (Coy, unpublished results) due to starting current transients.

The highly variable spatial properties of discharges between gapped area electrodes are known from a few studies that describe, variable, oscillatory, or fractal behavior [16-19]. There is a close relationship between the temporal and spatial variation due to the coupling between spatial path length, discharge current, ion production, and ion fragmentation.

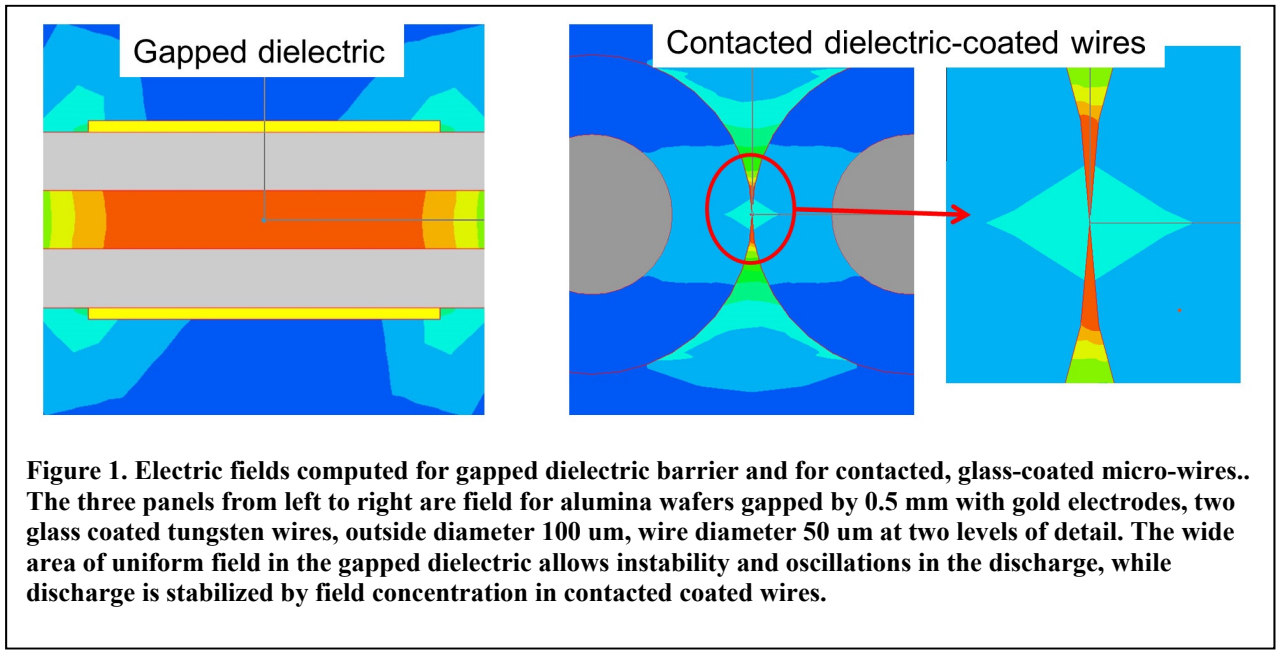

**Figure 1. Electric fields computed for gapped dielectric barrier and for contacted, glass-coated micro-wires.. The three panels from left to right are field for alumina wafers gapped by 0.5 mm with gold electrodes, two glass coated tungsten wires, outside diameter 100 um, wire diameter 50 um at two levels of detail. The wide area of uniform field in the gapped dielectric allows instability and oscillations in the discharge, while discharge is stabilized by field concentration in contacted coated wires.**

In order to quench the temporal and spatial discharge instability, we have considered an alternative design. The electric field magnitudes for gapped and contacted glass-coated wires are shown in Figure 1 (Maxwell SV, ANSYS). The left panel shows that the field within a 0.5 mm gap from back-excited alumina ceramics is uniform over a wide area. This uniformity allows filament initiation to occur at multiple and varying locations, and allows multiple filaments to interact, as described in Dong et al [20]. The middle panel and the zoomed image in the rightmost show that the contacted dielectric enhances the field near the contact point (glass 100 μm diameter, center tungsten wire 50 μm diameter, high resolution grid of 7000 elements). This geometry minimizes spatial and temporal variability, and limits the maximum ion energy in the plasma. When the wires are crossed at an angle, the area contact is a saddle surface. A photomicrograph of the structure is shown in Figure 2.

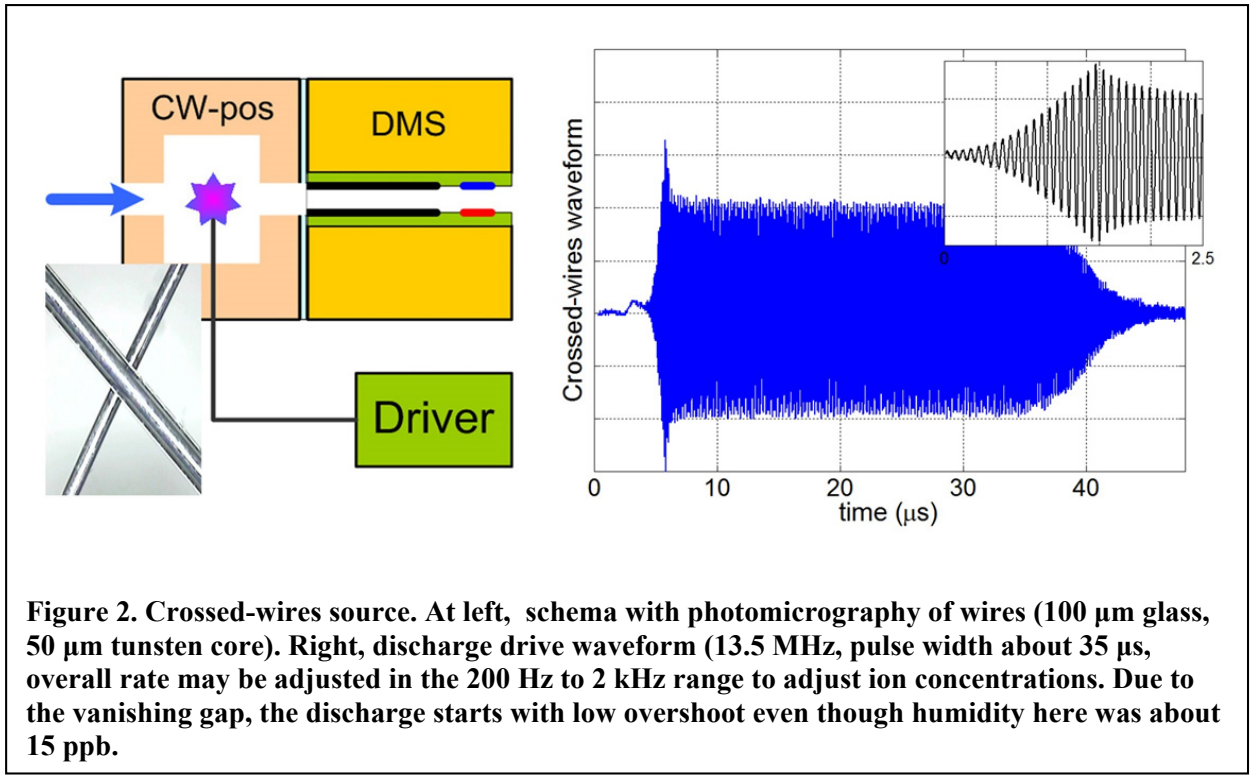

**Figure 2. Crossed-wires source. At left, schema with photomicrography of wires (100 μm glass, 50 μm tunsten core). Right, discharge drive waveform (13.5 MHz, pulse width about 35 μs, overall rate may be adjusted in the 200 Hz to 2 kHz range to adjust ion concentrations. Due to the vanishing gap, the discharge starts with low overshoot even though humidity here was about 15 ppb.**

The crossed-wires source operates in pulsed mode with little starting over-voltage, as shown in Figure 2. This results in pulsed ion production with an average ion flux that depends on the repetition rate of the drive pulse. The ion distribution is smoothed due to diffusion and pulses can be eliminated by detection synchronous with the source pulse rate or by averaging. The flow / clearance rate and the detection averaging can be matched to the ion source repetition rate to avoid on-peak oscillation at the source repetition rate.

## Experimental methods

### Source - Sensor Combinations

The crossed-wires source and DMS detector used for testing of chemical response and ionization efficiency were constructed for an NIEHS (National Institute for Environmental Health and Safety) - SBIR (Small Business Innovative Research) funded program to develop instrumentation for rapid detection of volatile organic compounds (R43-ES021126-01). The DMS filter section consisted of glass composite circuit boards with DMS filter electrodes 10 mm length by 1.62 mm width, spaced by Teflon of 0.5 mm thickness. The ionization discharge driver is a resonant circuit controlled by a timer, allowing pulsed, variable ion generation at overall repetition rate from 200 Hz to 2 kHz (operated here at 300 Hz). The generating waveforms are shown in Figure 2, and schematics are provided in supplementary material. The separation voltage waveform for the DMS ion filter is of the flyback type described in Krylov et al. [21], and the Faraday cup detector current amplifier and detection electronics are those of a Sionex SVAC unit (this technology has been transferred to Draper Laboratories and Danaher). The bandwidth of the detection electronics is approximately 800 Hz, and the transconductance gain is 160 pA/V. The separation voltage (SV) in these systems is a repetitive asymmetry waveform applied across the 1.62 mm width of thea 0.5 mm gap to the other DMS channel surface at ground. The SV mean-to-peak amplitude is settable in the range from 500 V to 1500 V or may be zero, and is combined with the DC compensation voltage (COV) which may be constant or scanned by steps. Measured noise properties of the electronics are discussed in the Results section. More details about the DMS detector section can be found in several publications [5, 22, 23].

For measurement of ion source noise properties, three other systems were used in order to compare differing technologies. These systems used, respectively, a 5 mCi Ni-63 source, a 0.5 mm gapped dielectric discharge, and a second glass-coated crossed wires ionizer. The wires ionizer - DMS used for noise measurements employed an alternative set of detection electronics that made use of a digital IIR-filter (low-pass, with feedback), that reduced electronic noise by 5-10 dB, while reducing the bandwidth to about 200 Hz. These slower electronics have a lower tranconductance gain of 100 pA/V, but we report all noise results in terms of a ratio to power at the signal peak so comparisons are not affected. Comparison of Ni-63 sensors with the two types of electronics showed substantially identical results for source properties after the electronic noise is removed.

### Noise measurements

Noise analysis of each of the units was based on high sample rate recordings of the signal from the positive reactant ion peak (RIP) in dry nitrogen. The gapped discharge and Ni-63 systems were recorded as repeated groups of 244 acquisitions, each averaged over a 1 msec gate time. A 1 msec time delay was required between the groups. Noise power spectra were computed using multiple varying length segments extracted from the data stream without spanning groups. The crossed wires source used a set of electronics which required a minimum of 5 msec averaging and 5 msec delay between groups. The longer acquisition time limited the maximum frequency which could be analyzed, however, the lowest frequencies are of the greatest interest because they have the greatest impact on speed of data acquisition.

The ion signal with electronic noise of the Faraday plate detector is observed with SV zero, and the electronic noise alone is observed with the DMS compensation voltage blocking transmission of all ions. In DMS transparent mode, both separation voltage (SV) and compensation voltages (COV) are zero allowing all ions to pass; in blocking mode, SV is zero, but COV is set outside of the ion transmission peak width (COV +5 V was used.).

## Chemical testing

Chemical tests on the crossed-wires system were performed in two modes, flow injection, and gas chromatography (GC) of diluted liquid samples, as schematically illustrated in Figure 3. Measurements were made with the sensor and gas flows near room temperature, in the range from 30°C to 35°C.

1. **Flow injection.** A constant flow of 0.85 to 1.0 slpm dry nitrogen (humidity 12 ppmv to 18 ppmv) was used as the transport gas through the ionizer and through the DMS filter and detector. Chemical vapor from a Hamilton gas-tight syringe of 10 µL capacity in a Harvard syringe pump was merged into the transport flow. In order to prevent trace contamination, the flow was first set with the needle valve and a flow meter, but

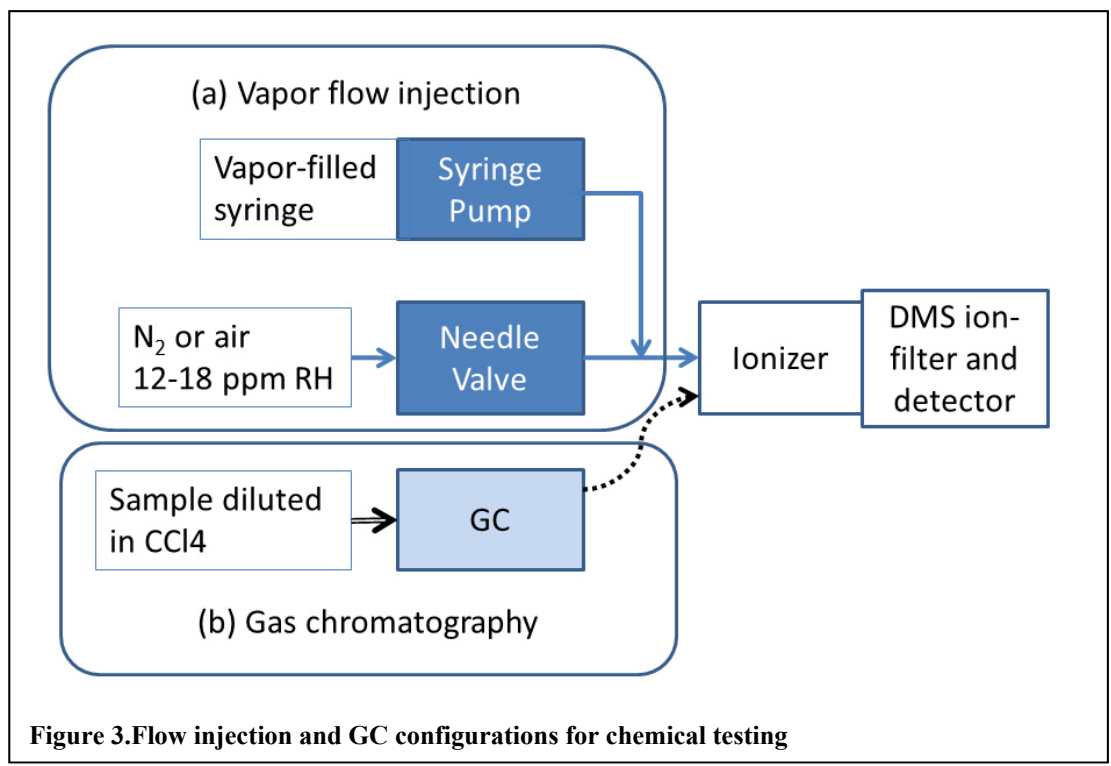

**Figure 3. Flow injection and GC configurations for chemical testing**

measurements were made with flow meter removed. Chemical vapor were loaded into the syringe from headspace of lab supply bottles of pure compounds. Stainless steel tubing and fittings were used. Vapor flows ranged from 0.25 μL/hour to 10 μL/hour, providing a highly dilute sample whose concentration could be varied using the syringe pump settings. A two dimensional scan of both SV and COV was recorded at multiple flow rates, allowing estimation of sensitivity and linearity

2. **GC delivery of diluted samples.** In positive ion mode, the wires source responds weakly to carbon tetrachloride ($CCl_4$), enabling its use as a solvent for serial dilution prior to GC injection of 0.5 or 1 μL. A GC split of 30:1 was also used in many cases even after high dilution to reduce concentration to non-saturating levels. The GC output flow was merged with the transport flow delivered to ionizers and DMS detector. The GC method allows detection of compounds in a complex mixture, prevents competitive ionization effects and can allow a measurement of relative ionization efficiencies. In practice, it was found difficult to achieve sufficient dilution for linear response unless peaks were tailed. Multiple species were easily detected, as the system has no significant memory effect.

## Results

### Noise analysis

A comparative frequency-resolved power spectrum of the signals from each of the systems record on the ion transmission peak, and on the ion-free baseline are shown in Fig 4 (a,b,c) for systems using the crossed glass-coated wires ionizer, the gapped dielectric barrier discharge ionizer, and the standard 5 mCi Ni-63 ionizer. The "On-peak" curves include ion source noise and detector noise, while "Baseline" curves include only electronics noise because the DMS compensation voltage is set to 5 V across the 0.5 mm gap to prevent ions from reaching the detector without being neutralized on the gold-metallized electrodes. All data was taken with separation voltage (RF amplitude) at zero. Noise is expressed as noise power, $P_{dB/Hz}$, in dB/Hz, $P_{dB/Hz} = 10\ log_{10}(P/P_0)$. The reference power, $P_0$, is set to peak-normalize each system separately, so that a constant value at the mean peak height yields 0 dB/Hz at the zero of frequency, and a pure signal oscillating from zero to the full peak intensity would also result in 0 dB/Hz. Baseline values show only electronic noise while the on-peak traces include electronics and source noise. Recordings were 3 to 5 minutes in length.

The power spectral density curves were computed using the MATLAB Signal Processing toolbox and the Welch method, using a Chebyshev window with 60 dB side lobe attenuation. The entire recording of on-peak or baseline data was treated as a single segment.

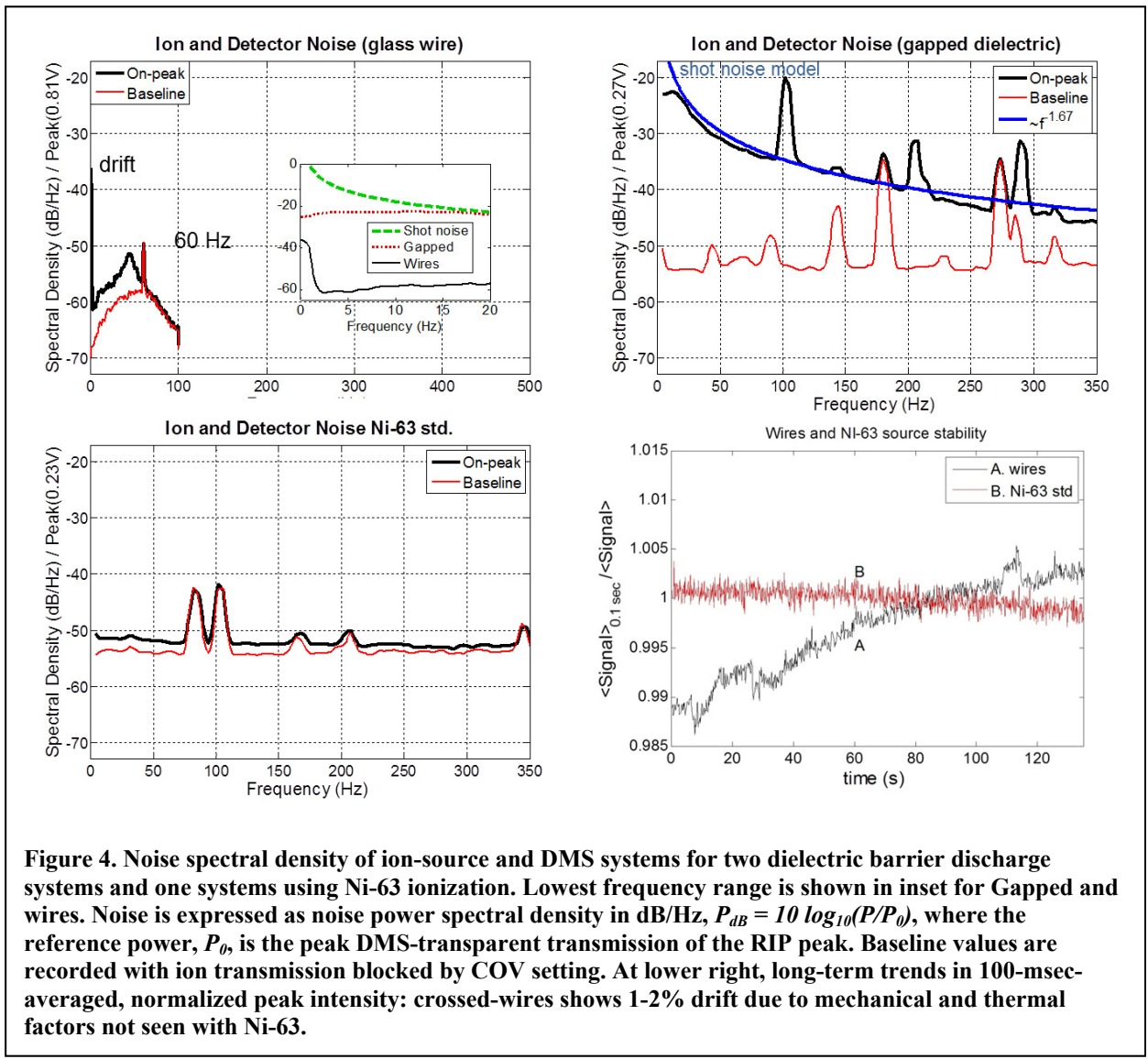

**Figure 4.** Noise spectral density of ion-source and DMS systems for two dielectric barrier discharge systems and one systems using Ni-63 ionization. Lowest frequency range is shown in inset for Gapped and wires. Noise is expressed as noise power spectral density in dB/Hz, $P_{dB} = 10\ log_{10}(P/P_0)$, where the reference power, $P_0$, is the peak DMS-transparent transmission of the RIP peak. Baseline values are recorded with ion transmission blocked by COV setting. At lower right, long-term trends in 100-msec-averaged, normalized peak intensity: crossed-wires shows 1-2% drift due to mechanical and thermal factors not seen with Ni-63.

The glass-coated-wire noise power spectrum, with the scale normalized to the RIP peak intensity, is shown in the upper left of Figure 4 along with an inset that compares the lowest frequency range of the glass-coated wires and the gapped dielectric. The baseline trend is characteristic of the IIR-filter-enhanced electronics (5 ms. step) used with this ionizer. Except for the small 60 Hz peak (AC line frequency interference), the noise level is low and featureless except very close to zero frequency where a significant peak occurs, which is expanded in the inset figure. The Figure 4 lower left panel from the 5 mCi Ni-63 ion source with the standard electronics (1 ms. step). The Ni-63 noise levels in the lower left panel are frequency-independent (white noise), with little difference in noise level between the peak and baseline data. Ni-63 peak and baseline noise levels further approach each other at higher frequencies. There are a few small peaks present in both peak and baseline which are believed to be due to power supply noise. The glass wire system produces a higher ion flux than the Ni-63 source, so the peak-normalized, relative noise levels are lower.

The lower right panel of Figure 4 reveals the drift in the wires source by comparing peak signals averaged over 100 ms. intervals for Ni-63 and glass wires. The wires source limits the accuracy of peak intensities to 1-2% of the peak intensity. Improving the rigidity of the support for the crossed wires seen in Figure 2 would be expected to improve this performance, although an accuracy of a few percent is more than is required in applications, due to the presence of other influences on intensity such as modifiers, including humidity level [24-28]. With the wires source, a 5-10 msec. averaging time is sufficient to read a 1% accuracy level for most compounds; the source noise is proportional to ion current, (approx. 0.5% of peak (1 sigma)) while the electronic noise is (approx. 2-3 mV, (1 sigma)). The enhanced digital filter of infinite impulse response (IIR) type in the wires source electronics includes feedback that also contributes in the low frequency region.

To continue the discussion of noise sources, we examine the characteristics of the transconductance amplifier. The baseline noise of the standard electronics, from the Ni-63 baseline level, may be compared with the noise properties of commercial a picoammeters. The transconductance gain is 160 picoamp per volt (pA/V). At an averaging time of 100 msec, omitting the RIP peak normalization of Figure 4, the RMS noise is 0.147 mV RMS. Referred to the input, this then corresponds to 23.5 fA noise current. The noise specification for the Kiethley 6487 picoammeter is 20 fA at the input at low frequency with the same acquisition repetition rate, 10 Hz. Thus, our observed performance is consistent with optimized technology. We can also note that the gain corresponds to is $10^6$ charges per second per millivolt (1 pA is 6.2415 x $10^6$ charges per second), which can provide a comparison to counting statistics. Excess noise at low frequencies typically present in systems with response to DC, and it can often be addressed by the use of phase-sensitive detection or modulation techniques in order to suppress very low frequency noise or drift, and it can be used to reduce local buildup of charges that can occur in DMS systems [22].

The gapped dielectric noise seen in Figure 4(upper right) has a median baseline similar to the Ni-63 standard system, but also shows numerous additional peaks in both the baseline and in the ion power spectral density. Peaks common to the peak and baseline are due to electrical pickup from the high peak currents in the discharge filaments between electrodes during each cycle of the discharge waveform. The discharge is an area discharge with filaments across a 0.5 mm gap that is only two centimeters in direct line from the Faraday plate detector. Area discharges show cooperative oscillations [20] between multiple discharge locations that can also contribute to periodic peak and baseline. The on-peak ion current has additional noise that increases at low frequency, and appears as a rising baseline which is approximated by a $(1/f)^{1.67}$ curve. This type of noise is variously known as 1/f, shot, or flicker noise (see the fine tutorial review by Milotti [29]), and is well-known in many contexts [30]. The power, $\alpha$, in the $(1/f)^\alpha$ dependence ranges from 0.5 to 2 in different physical systems. The systems with $(1/f)^2$ behavior can be interpreted in terms of a Brownian motion kernel, which in this case would originate from random motion of the discharge location and intensity with the current reversals and between the 13.5 MHz tone bursts. Movement of the discharge location is limited at the highest frequencies by local heating and the persistence of charge density along the discharge path and so does not continue to diverge at the lowest frequencies (inset in Figure 4 upper left). Shot noise is harmful for quantitative analytical applications because noise at low frequencies is the difficult to remove by time-averaging.

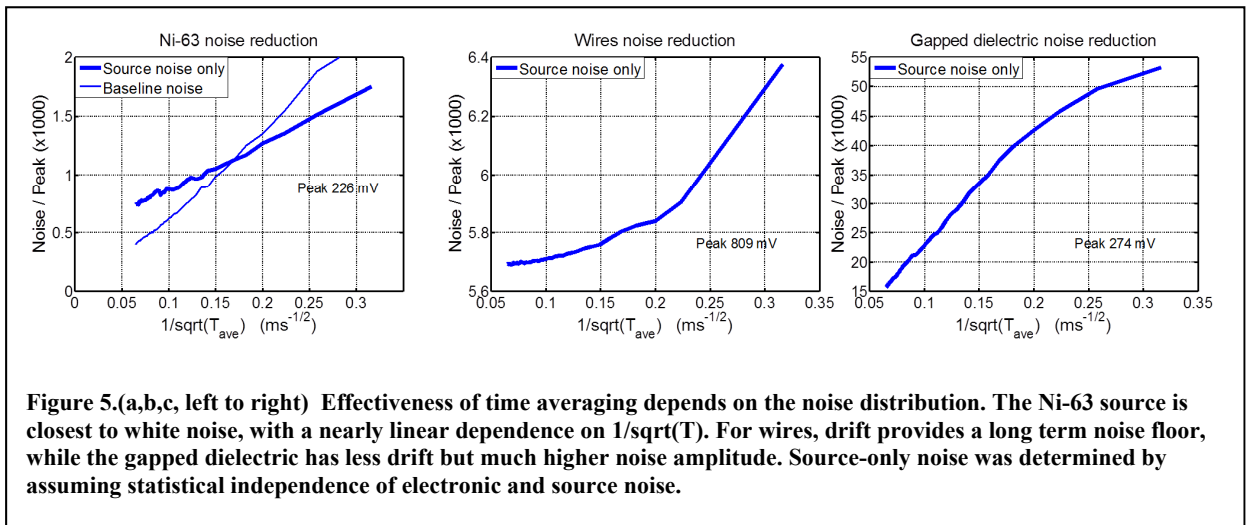

**Figure 5.(a,b,c, left to right)** Effectiveness of time averaging depends on the noise distribution. The Ni-63 source is closest to white noise, with a nearly linear dependence on 1/sqrt(T). For wires, drift provides a long term noise floor, while the gapped dielectric has less drift but much higher noise amplitude. Source-only noise was determined by assuming statistical independence of electronic and source noise.

The effect of low frequency noise on residual source noise from time averaged signals is shown in Figure 5 for three systems: Ni-63, crossed wires source, and gapped dielectric. The excess source noise is computed for the noise on peak by assuming peak and source noise are uncorrelated, so that variances of source and electronic noise are additive. With a white noise source the variance of the random noise is inversely proportional to the averaging time, so we have plotted in the three panels of Figure 5 the source noise standard deviation (electronic noise removed) against $(1/T_{ave})^{1/2}$. Because the Ni-63 baseline noise shown on the Ni-63 panel is linear and tending to a near-zero intercept, electronic noise is essentially white. The Ni-63 source noise is linear, but with a non-zero intercept, indicating an additional low frequency contribution such as unsteady flow or thermal effects giving at the weak long term drift seen in Figure 4. The wires source has a greater drift contribution, so the limiting noise level is greater than for Ni-63. Part of this plateau in the wires system has been found to be due to the feedback in the IIR digital filter implemented in the electronics: when comparison is made to Ni-63 with the more complex electronics, a plateau is reached at a level of about 3.6 on the above scale. The gapped dielectric has a still higher noise level, but, at longer averaging times, appears to be decreasing linearly with $T^{(-1/2)}$. If speed is sacrificed, it appears that the gapped dielectric could be made competitive by using longer averaging times on known targets, although the inset in Figure 4 shows that it remain higher than the crossed wired. A key advantage of DMS detection paired with the crossed wires source is the combination of high speed, and sufficiently low noise, reaching 1% readings in a few ms.

The noise analysis shows that vanishing gap of the crossed-wires geometry stabilizes the discharge location and reduces the starting current, resulting in near-absence of (1/f) noise. The geometry also allows the starting voltage to be essentially independent of humidity level, with the system showing stable operation even at low ppb humidity levels.

**Chemical response**

Tests were conducted using differential mobility spectrometry (DMS) detectors operating in $N_2$ with humidity levels between 12 ppmv and 325 ppmv at the Las Cruces, NM, atmospheric pressure of 12.7 psi (87.56 kPa), with sample supplied by vapor flow injection or gas chromatography. The tested organics include a wide range of aliphatics, aromatics, alcohols, and ketones. The non-radioactive glass-wire source is able to ionize aliphatics such hexane and octane in its wedge-gap discharge. These compounds are difficult to detect with Ni-63, which ionizes largely by proton-transfer through a water reactant ion peak, with a minor contributions from direct beta ionization (see diphenylmethane cation in Kendler et al. [23]), and from other minor species. This extended chemical coverage is characteristic of discharge sources, and has also been seen with glow discharge (for instance, Dong et al. [31]).

*Protonated monomers and proton-bound dimers*

As an example of higher proton affinity analytes, flow-injection data for 1-pentanot is shown in Figure 6. Vapor collected from the headspace of 1-pentanol liquid in a Hamilton gas-tight 1 μL syringe was merged with 1.0 SLPM flow of nitrogen containing 17.5 ppmv of moisture at a temperature of 30 to 35 °C with the system operated local atmospheric pressure of 87.56 kPa (12.7 psi). Under these conditions, corresponding to 20.3 ppbv and 40.6 ppbv of 1-pentanol, the reactant ion peak is seen to be partially depleted, and both protonated monomer and proton-bound dimer are observed. The DMS voltages are applied across the 0.5 mm gap of the DMS filter region, and the separation voltage (SV) (of the flyback type [21]) is stepped in mean-to-peak amplitude from 500 V to 1400 V in steps of 20 V, while the compensation voltage (COV) is scanned from -40 V to + 5 V in steps of 0.2 V, averaging 10 ms. at each point. In the left panel, a topographic plot shows that a transition is observed from the dimer or cluster peak to the monomer peak with increasing separation voltage. Monomer-dimer and de-clustering kinetics are evident here, and they have been studied in a number of publications, including studies with mass-spectrometric resolution of the species involved [32-35]. Mass spectrometric results and scaling arguments allow the 1-pentanol peak at more negative COV (middle features at left, and -7 V in right panel) to be identified as the monomer, additionally neutral-clustered below 800 V SV.

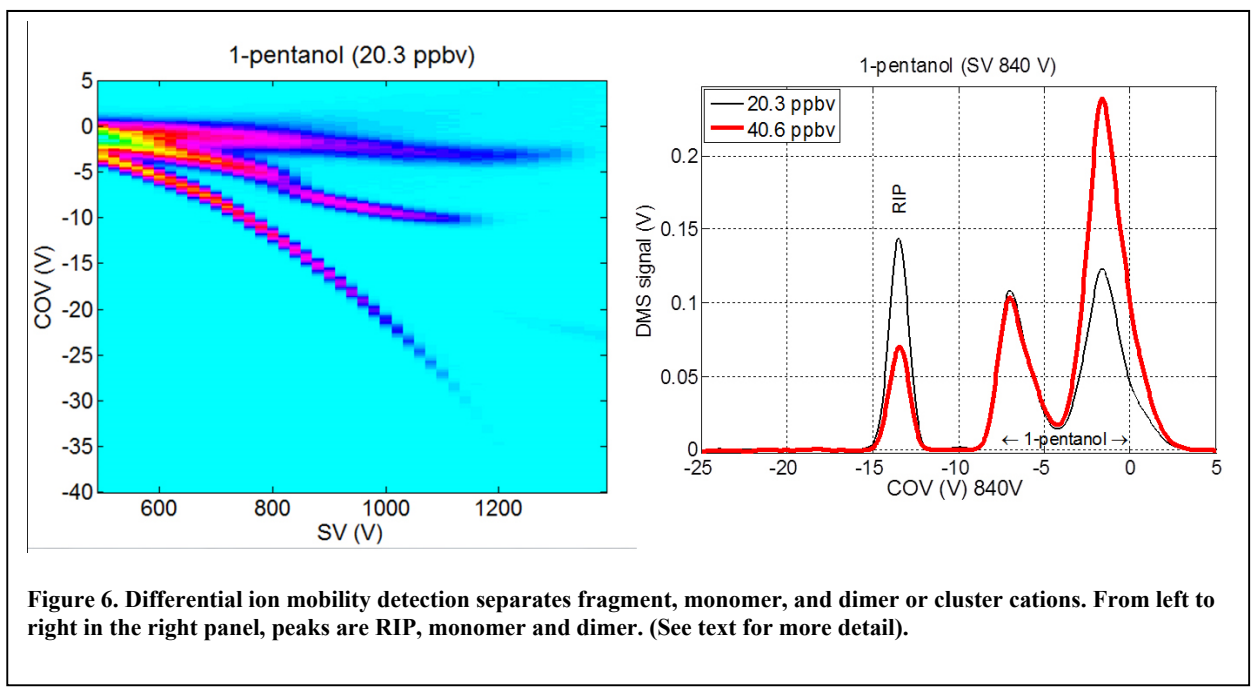

**Figure 6. Differential ion mobility detection separates fragment, monomer, and dimer or cluster cations. From left to right in the right panel, peaks are RIP, monomer and dimer. (See text for more detail).**

      Estimates of the efficiency with which neutral species are ionized and detected, are affected by clustering with neutrals and ions. These effects depend on bulk gas temperature, and on pressure, and can be minimized or suppressed by increasing temperature either pre- or post-ionization [23, 25, 33, 36-39]. For an approximate and conservative efficiency measurement, we choose a separation voltage which isolates RIP and individual chemical peaks, select one of the responding peaks and use it to estimate ionization efficiency.  In this case the monomer peak alone at 940 V is selected (a higher value than is shown in the right panel), where the monomer peak has grown from DMS-heating-induced dimer dissociation [23] and has become larger than the dimer at the 20.3 ppbv concentration.

## NO$^+$ clusters: Benzene, Toluene

The topographic plot for benzene at a humidity level of 17 ppmv and the response curve as a function of concentration for benzene are shown in Figure 7. Benzene is not detectable with Ni-63 at humidity levels above the mid-ppb levels. It has a relatively low proton affinity of 750 kJ/mol but is more difficult to detect with radioactive ionization than might be expected from that value alone. Similarly toluene (PA 784 kJ/mol) is not detectable with Ni-63 at humidity levels above 2-5 ppmv. The signal level for the benzene peak at SV 800V is plotted in the right panel, and leads to a threshold sensitivity of 55 ppbv-slpm/mV. In the benzene topographic plot there is a transition at about 950 V to a much weaker feature with the DMS field. In the temperature range used for these measurements, 35±3°C, toluene has been found in accurate mass time of flight mass spectrometry to cluster with NO$^+$. A large collection of NO$^+$ affinities [40] gives a NO$^+$ binding energy of 130.96 kJ/mol for benzene, and 144.3 kJ/mol for toluene. Aromatics interact with π electrons with NO$^+$, but σ bonds with H$^+$. NO$^+$ affinities and proton affinities are correlated in two groups: compounds with aromatic rings the interact with NO$^+$ through πelectrons follow one trend, and compounds interact through σ interactions follow a separate linear trend [40]. The NO$^+$ concentration from the wires ionizer is significantly higher than with Ni-63, leading to detection as M·NO$^+$ as well as, or instead of detection as M·H$^+$.

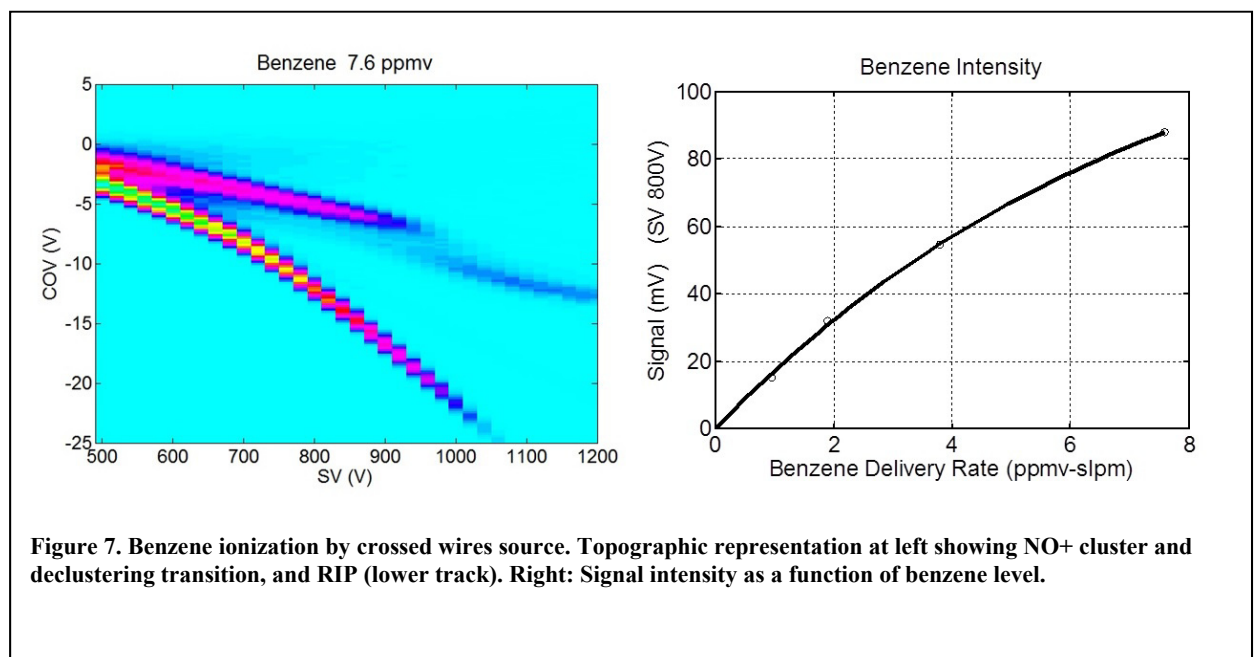

**Figure 7. Benzene ionization by crossed wires source. Topographic representation at left showing NO+ cluster and declustering transition, and RIP (lower track). Right: Signal intensity as a function of benzene level.**

Mass spectrometry of Ni-63 RIP species, even in sub-ppm humidity air or nitrogen, consistently show much higher concentrations of protonated water dimer (PA 808 kJ/mol [41]) than hydronium cation (PA 691 kJ/mol) except with low ppb humidity and significant heating due to DMS field (unpublished results). This experimental observation, and the lack of fragmentation of benzene and toluene in DMS with either Ni-63 or wires discharge ionization, can account for the enhanced, but relatively low sensitivity to these compound (see Discussion section). The relatively soft fragmentation of the wires discharge source leads to detection of additional analytes.

*Fragmentation*

Several of the tested compounds show additional peaks in positive ion mode due to fragmentation that are not seen with the Ni-63 source, broadening the chemical coverage. Among these are ethylbenzene, which shows both a fragment peak and a peak that is interpreted as $\pi$-clustered $NO^+$. Fragment peaks are not seen for benzene or toluene, and the ethylbenzene fragment cation leads to improved sensitivity.

As another example, chloroform, $CCl_3H$, and carbon tetrachloride, $CCl_4$, generate two cations which appear with different intensities for the two compounds, as shown in Figure 8. The differential mobility peak labeled (1) appears with a higher intensity in chloroform than in carbon tetrachloride, while the relative intensities are reversed for peak (2). As peak (1) appears at a more negative compensation voltage, and has a smaller linewidth than (2), general properties of differential mobility indicate that peak (1) is the fragment with a smaller geometric size / molecular weight, and that (2) is a larger ion, but still a fragment since it appears for both molecules. These two compounds are not seen with significant intensity with Ni-63.

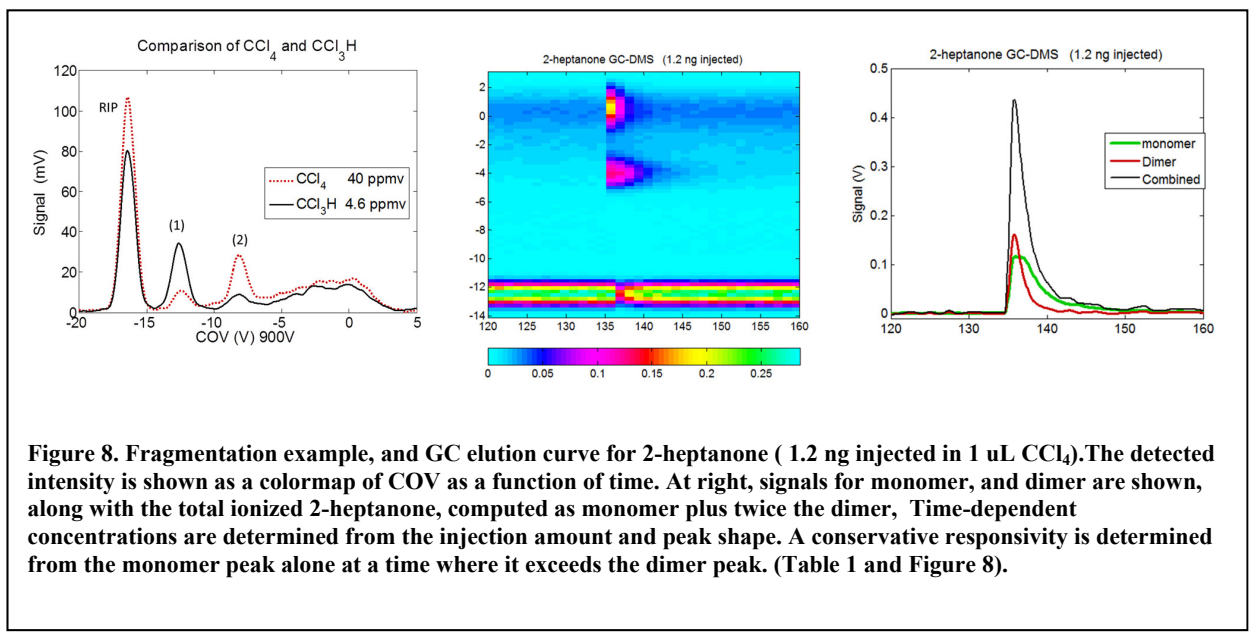

**Figure 8. Fragmentation example, and GC elution curve for 2-heptanone ( 1.2 ng injected in 1 uL CCl₄).The detected intensity is shown as a colormap of COV as a function of time. At right, signals for monomer, and dimer are shown, along with the total ionized 2-heptanone, computed as monomer plus twice the dimer, Time-dependent concentrations are determined from the injection amount and peak shape. A conservative responsivity is determined from the monomer peak alone at a time where it exceeds the dimer peak. (Table 1 and Figure 8).**

*Sample Introduction by GC*

GC-DMS data was collected for a number of mixtures and individual compounds, including aryl compounds (benzene, toluene, ethyl benzene, propyl benzene, xylenes), alcohols (methanol, isopropanol, 1-pentanol, 1-hexanol), aliphatics (hexanes, n-octane), and ketones (2-pentanone, 2-heptanone). A 15 m DB-5 column was used, with nitrogen carrier gas, with temperature profiles adjusted according to the mixture, but with column temperature remaining below 50°C. Serial dilutions were made in $CCl_4$ for 0.5 or 1.0 µL GC injections. The GC output was coupled to the ionizer and DMS through a 30:1 split in most cases in addition to dilution. The GC data was useful in testing for memory effects, suppressing baseline contributions, and detecting mixtures without competitive ionization, but was not generally used for responsivity calculations because concentrations outside the linear range were reached in the eluted peaks.

The ketone group of compounds is among the most efficiently ionized, followed by the alcohols. Figure 8 shows section of the GC data for 2-heptanone, under conditions that permit the responsivity to be determined from an analysis of the peak shape. In the GC data for 2-heptanone, 1.2 ng was injected. In Figure 8(center), the detected intensity is shown as a colormap of transmitted ion signal at COV values as a function of time from 120 sec to 160 sec. At right, signals for monomer, and dimer are shown, along with a curve for the total ionized 2-heptanone, computed as monomer plus twice the dimer, showing a near-exponential decay of total concentration. Time-dependent concentrations are determined from the injection amount and combined peak shape. A conservative responsivity is determined from the monomer peak alone at a time where it exceeds the dimer peak (see Summary of results, Table 1 and Figure 8).

*Summary of results*

The chemical compounds for which approximate responsivity levels have been determined are listed in Table 1. These values were determined from flow-injection data such as that shown in Figure 6 and 7, with the exception of 2-heptanone where the tailing of the monomer GC peak was used. The ionization efficiency varies with functional groups, with ketones and alcohols being most easily detected, followed by aromatics, and aliphatics. Additional compounds which were observed are shown in Table 2.

Responsivity or level of detection is reported as sample flux (ppbv-slpm) per mV of detected signal in the wires-ionizer – DMS system. Detector transconductance of 160 pA/volt is equivalent to $10^6$ charges /sec/mV. As 1 ppbv-SLPM is $4.419 \times 10^{11}$ molec/s, the overall sample-to-detection efficiency is $2.26 \cdot 10^{-6}$. Ionization efficiency can be increased by increasing the duty factor of the wires-ionizer drive voltage above the 300 Hz used for these test.

| Sample | Proton Affinity | ppbv/mV at 1 SLPM |
|---|---|---|
| 2-heptanone | 845 | 0.025 |
| 2-pentanone | 833 | 0.035 |
| 1-pentanol | 795 | 0.141 |
| ethylbenzene | 788 | 3.43 |
| toluene | 784 | 60.3 |
| nitromethane | 755 | 1.71 |
| benzene | 750 | 54.8 |
| diesel (PA est.) | 750 | 1.00 |
| hexanes | 684 | 74.0 |
| octane | 675 | 33.5 |
| trichloroethylene | 670 | 30.5 |
| CHCl3 | 659 | 335. |
| CCl4 (PA est.) | 650 | 702 |

Table 1. Ionization threshold sensitivity as sample flux (ppbv·slpm) per mV of detected signal in the wires-ionizer – DMS system. A value of 1 ppbv·slpm/mV is an overall sample-to-detection efficiency of $2.26 \cdot 10^{-6}$. Ionization efficiency can be increased by increasing the duty factor of the wires-ionizer drive voltage. (see Discussion section).

| Sample | Proton Affinity kJ/mol |
|---|---|
| benzaldehyde | 833.9 |
| acetone | 812.1 |
| 1-hexanol | 799.0 |
| xylenes | 794.5 |
| isopropanol | 792.4 |
| methyl ethyl ketone | 792.4 |
| propylbenzene | 790.0 |
| methanol | 754.3 |

Table 2. Other tested samples. These compounds were readily detected by GC injection alone and in mixtures at low levels consistent with those in Table 1.

## Discussion

The ion generation efficiency of the crossed, glass-coated wires ionizer varies with chemical species, and shows a dominant correlation with proton affinity. Figure 9 displays this trend by plotting, from Table 1, the sample delivery required for 1 mV signal as a function of proton affinity, using units for required sample level of ppbv·slpm for 1 mV signal. With 100 msec averaging, the RMS electronic noise is less than 0.6 mV (Figure 5a), so 1 sec averaging would result in under 0.2 mV noise, omitting other systematic effects and interferences.

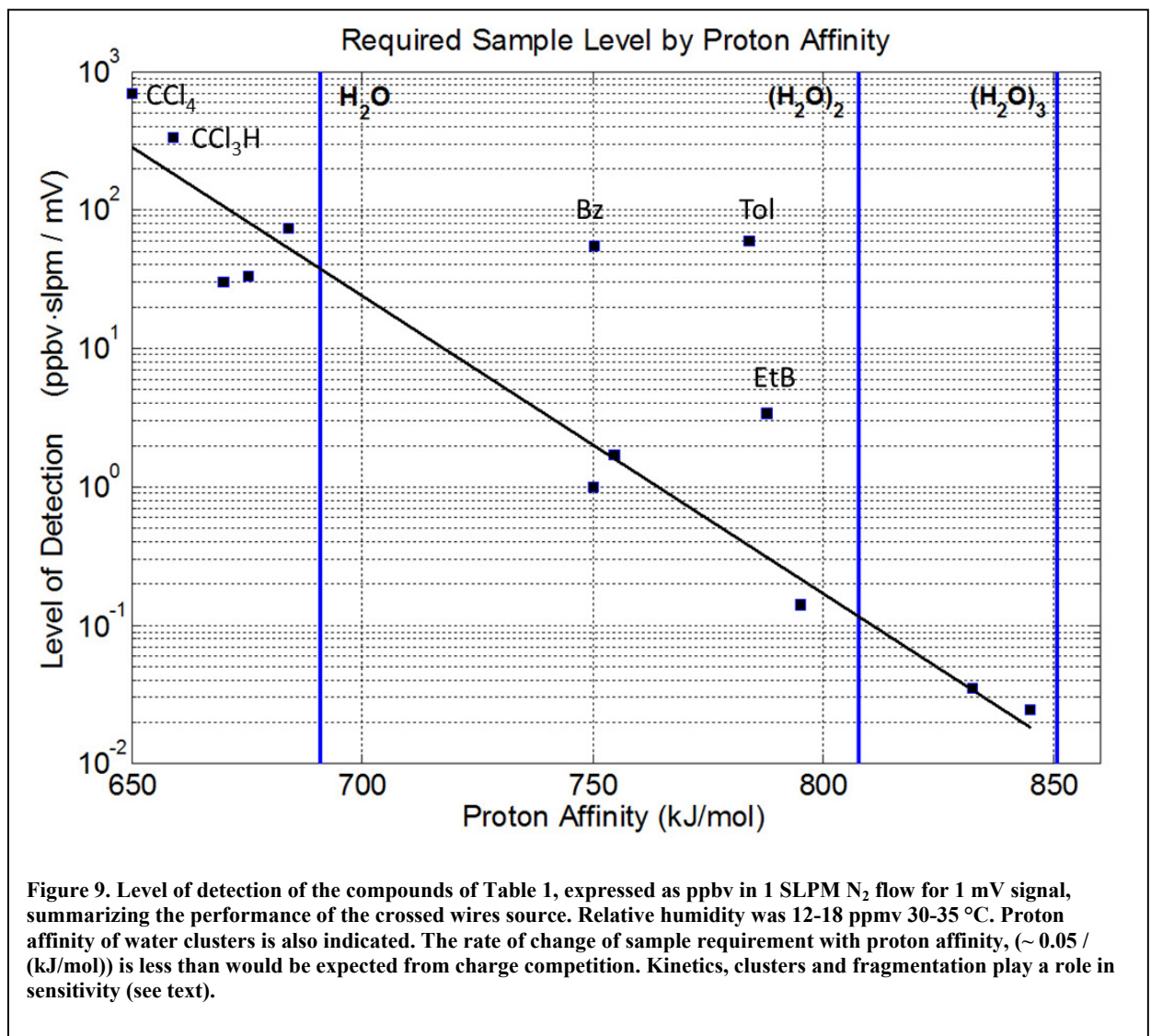

**Figure 9.** Level of detection of the compounds of Table 1, expressed as ppbv in 1 SLPM $N_2$ flow for 1 mV signal, summarizing the performance of the crossed wires source. Relative humidity was 12-18 ppmv 30-35 °C. Proton affinity of water clusters is also indicated. The rate of change of sample requirement with proton affinity, (~ 0.05 / (kJ/mol)) is less than would be expected from charge competition. Kinetics, clusters and fragmentation play a role in sensitivity (see text).

The bulk of the data on Figure 9 is near the inclined solid line that was generated by a robust linear fit the data. Robust fitting discriminates against outliers; here the MATLAB robustfit function was used with a Welch parameter for outlier suppression of 1.5. The slope of the line corresponds to a change in sensitivity of about 5% per kJ/mol; that is, the required sample level increases by about 5% for every kJ/mol decrease in proton affinity (PA). This is a slow rate of change of sensitivity; if the effect were based only on equilibrium barrier-less charge competition with one particular protonated compound, the rate of change would be (1/RT), or 40% per kJ/mol at 300 K. Several effects can be involved in this broadened chemical coverage. The creation of ionizable fragments is responsible for good sensitivity at low PA. Adducts of ions like $NO^+$ or other ions such as hydronium, contribute for aryl species and others. The charge balance between competing species is in an intermediate area between kinetic control and thermodynamic control because residence time between ion creation and detection is low. Kinetics and thermodynamics of ion-neutral reactions have been the subject of many studies and the chain of relevant reactions can be complex, and may include barriers that slow thermodynamic equilibrium [32, 42, 43]. The importance of kinetic control was critical in the determination of water cluster kinetics and and thermodynamics by the Kebarle group [44, 45]. Under the current conditions, the residence time of ions in the detection system is only 0.3 msec, and the sample concentrations are in the ppb range and the moisture level in the 15 ppm range, while the DMS waveform period is only 800 nsec [46]. Moreover, in the tested range of sample concentration, the hydronium (RIP) ion level is at approximately the same order as the sample ion level, so the collision rate between hydronium ions and neutral sample is low.

As we have mentioned, benzene and toluene are more difficult to detect than would be expected from their proton affinity, but both compounds are detected at higher moisture levels with the wires discharge than is seen with Ni-63. This sensitivity is believed to be related to an increase in clustering with $NO^+$. $NO^+$ concentrations are increased in the wires source, and $NO^+$ clustering with toluene is known from Ni-63 ionization of toluene DMS filtered, and detected by TOF mass spectrometry, where $Tol \cdot H^+$ was also observed at sub-ppm moisture levels (unpublished results). Ethylbenzene, though in the same class, is detected more efficiently due to the additional creation of a light fragment.

The wires source is capable of generating ion fluxes a factor of 10-30 greater than was employed in the current series of tests. The wires source was operated at 300 Hz (3.33 ms. period), with discharge bursts of approximately 40 μs (Figure 2). Since the gas flow resulted in a residence time (including dwell time in the ionizer region) of less than 0.5 ms., the repetition rate could be increased to 3 kHz or even 10 kHz to provide a more continuous ion stream. The operational settings were chosen to provide similar average ion loads to those generated by the other sources for compatibility with the detection electronics. Operation with a lower ion flow prevents static charge buildup in the miniature 0.5 mm x 1.62 mm x 10 mm sensor geometry [22].

The standard Ni-63 DMS system achieves detection sensitivities on the order of 30 ppt at 0.3 slpm or lower for S/N = 3, with 1 sec averaging time for strongly responding cations such as methyl salicylate (PA 850 kJ/mol) and dimethyl methyl phosphonate (PA 902. kJ/mol). Compounds with such high proton affinities were not in the current test set, but, converting to 0.1 sec averaging time, the 30 ppt value is equivalent to 0.028 ppbv·slpm/mV. This value is essentially the same as seen for the two 2-ketones, which have lower proton affinities. Overall performance thus equals or exceeds the performance of the system with radioactive ionization. Signal level from the wires ionizer can be increased by increasing the repetition rate of the discharge, allowing overall efficiency in the best case to go from tenths of a percent to a few percent for the complete system.

Using a conservative approach to operating and testing the crossed-wires ionizer, in comparison with a Ni-63 system, and a gapped dielectric-barrier discharge, results show the crossed-wires ionizer to be competitive with radioactive ionization, with speed, low noise, and efficiency capable of matching and exceeding the performance of other technologies. Limitations of the ionizer include the restriction to cations, and the complexity created by the creation of fragments. Adaptation of the source for negative ion generation should be possible following the designs that have been employed for IMS. Voltage extraction through an opposing gas flow is used to select high mobility negative ions so as to discriminate against larger high electron affinity species ($NO_3^-$, $CO_3^-$, $NO_2^-$) [47, 48]. The creation of fragments adds complexity, but is essential to obtaining sensitivity to species like alkanes and other types, and to their observation at significant humidity levels. The small discharge volume of the crossed wires source limits the degree of fragmentation to a lower level than can be achieved with a larger, more energetic, and electrically noisy design like a gapped dielectric barrier discharge.

## Acknowledgements

We gratefully acknowledge the support that has made this work possible. The development was initially supported by NIEHS SBIR R43-ES021126-01 (PI, S. Coy, Imaginative Technologies LLC). Additional support for the development and analysis of sensitive DMS methods was provided by R01 AI101798 (PI, Albert J. Fornace Jr., Georgetown U.).

## Competing Interests

The authors declare no competing financial interest.

Supplemental Information for "A gapless micro-dielectric-barrier-discharge ion source for analytical applications", Coy et al.
Crossed-Wires Ionizer Construction

# Contents

The **objective** of this project was development of a non-radioactive electric discharge ionizer for DMS. NRI-DMS is supposed to be used as a gas chromatographic detector.

## Implementation

Prototype consists of the discharger where the ionization takes place; electric driver providing necessary voltages, compartment containing the discharger, gas lines and electric connections, custom DMS sensor and SVAC electronics.

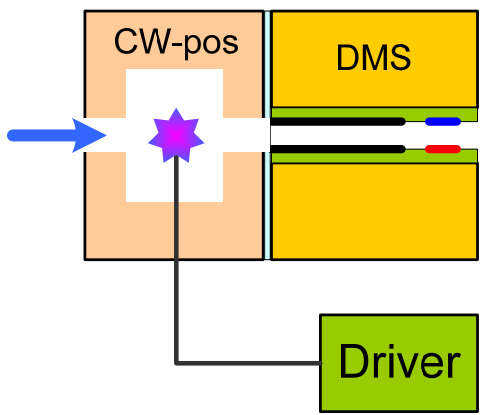

**Discharger**

Two crossed glass coated wires embody NRI discharger.

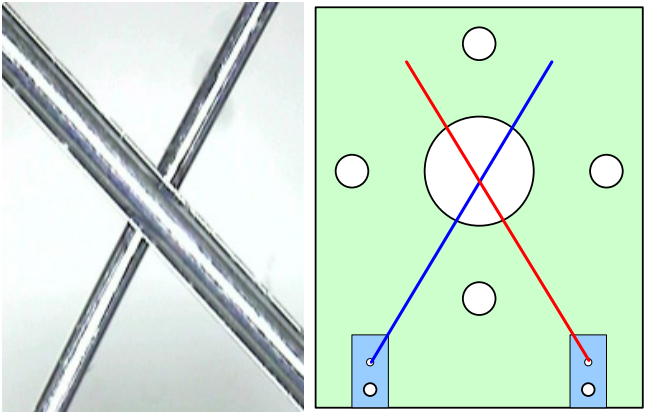

**Driver**

NRI driver provides necessary potentials and waveforms to initiate and maintain the discharge.

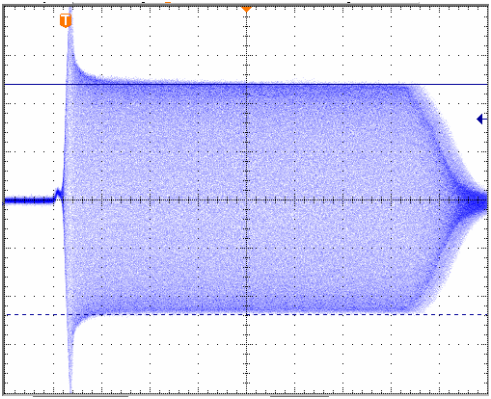

Block-scheme

Resonance generator working in a pulse mode with discharger as a component of the output *LC*-circuit was used.

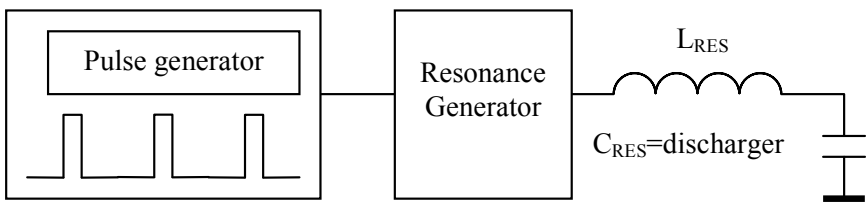

Schematics

Electrical circuit and PCB layout

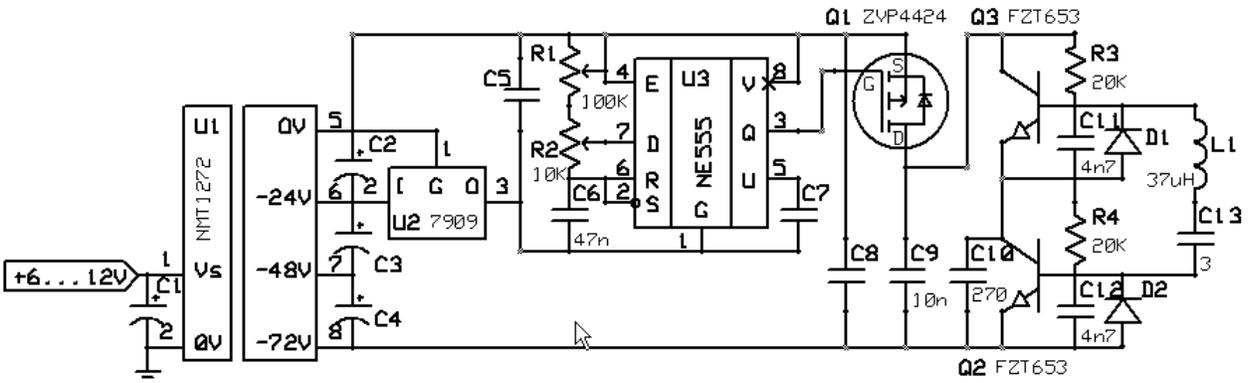

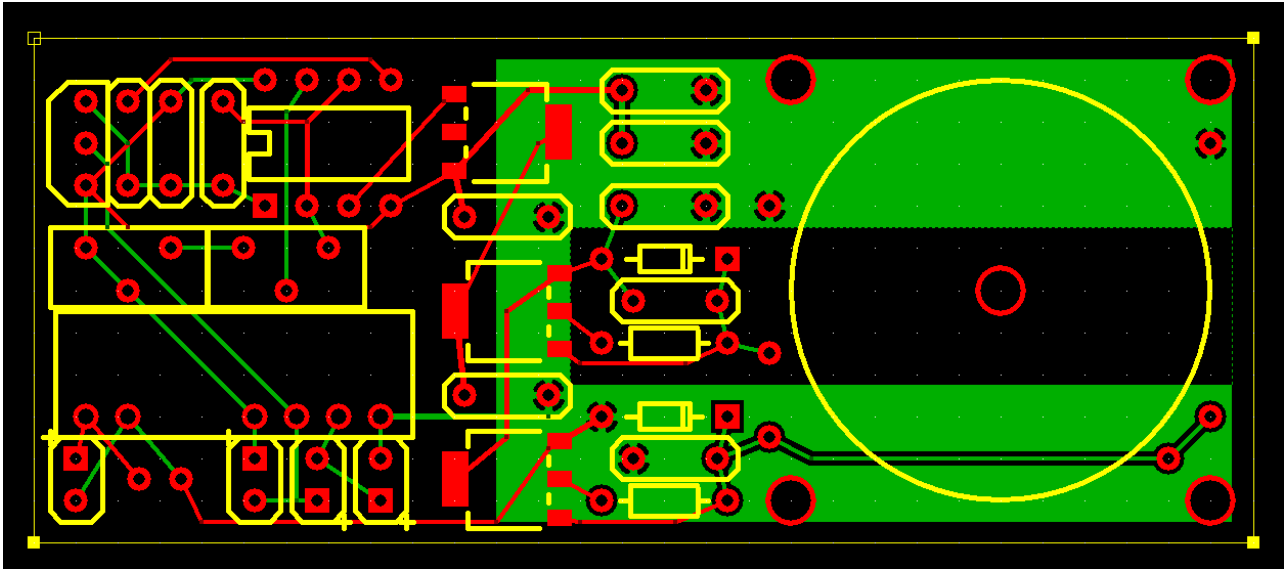

Technical Characteristics

- supply voltage 6 - 12 V;
- pulse repetition period 0.5 – 5 ms;
- pulse duty cycle 1 – 10 %;
- high frequency 13 MHz;
- power consumption 0.1…1.0 Wt.

**Compartment**

NRI compartment mechanically supports the discharger, provides necessary gas flows and electrical connections.

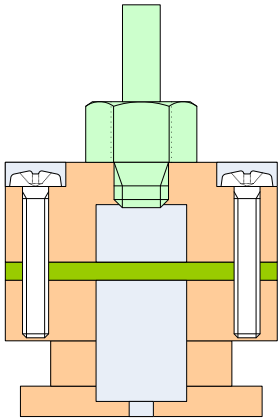

Lower part

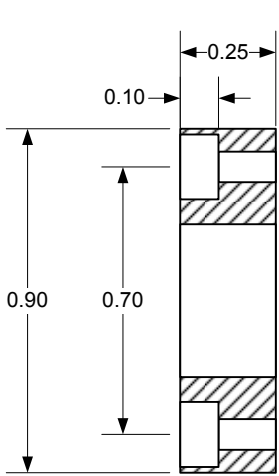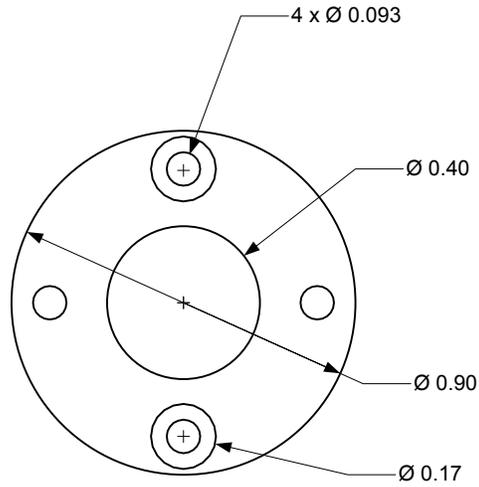

Upper part

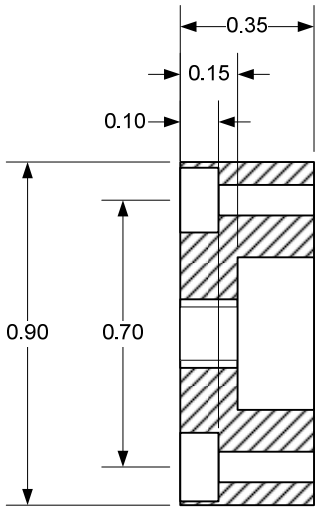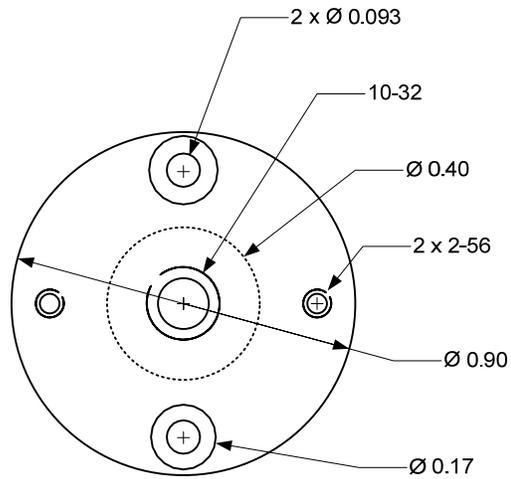

**DMS sensor**

Modified DMS sensor used with NRI consists of two plates, spacer, and manifold

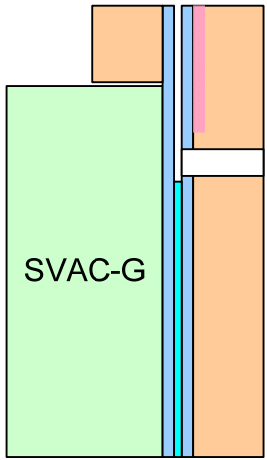
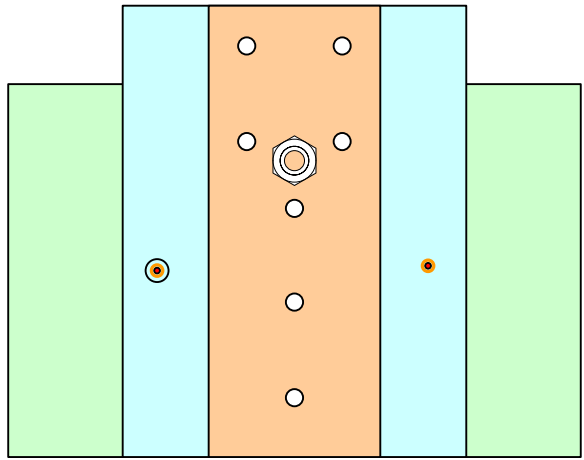

**Plates**

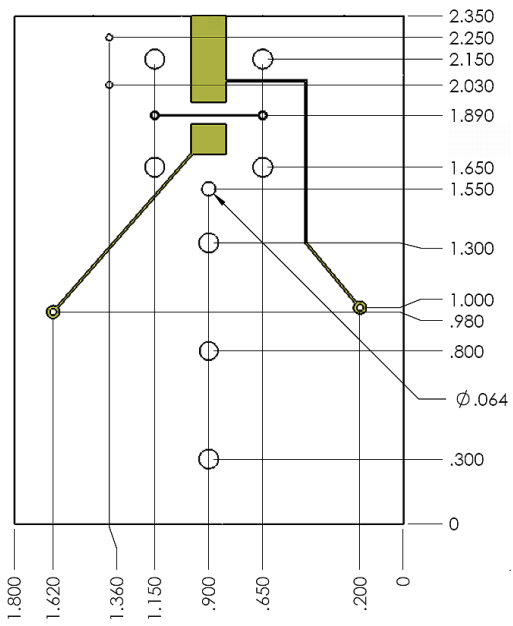
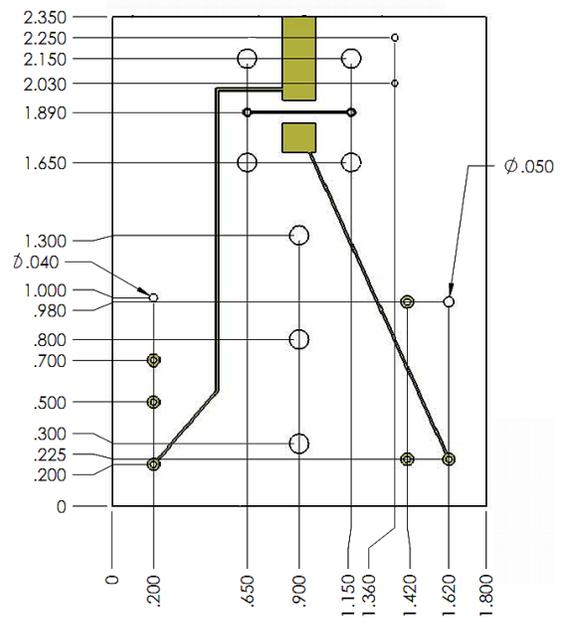

## Spacer

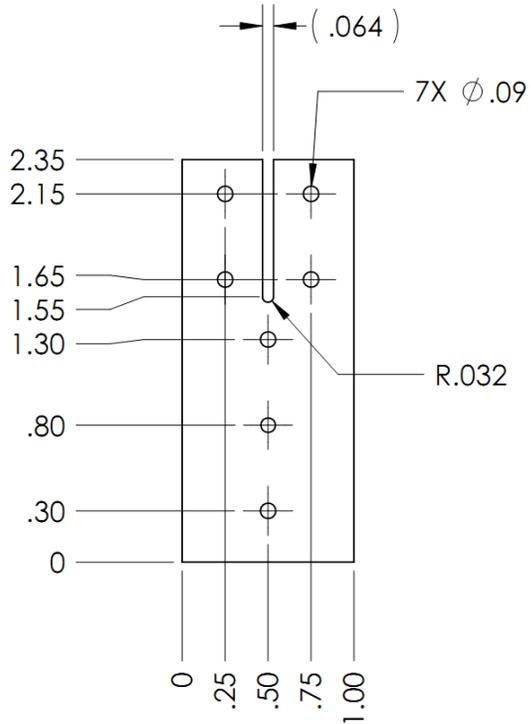

## Manifold

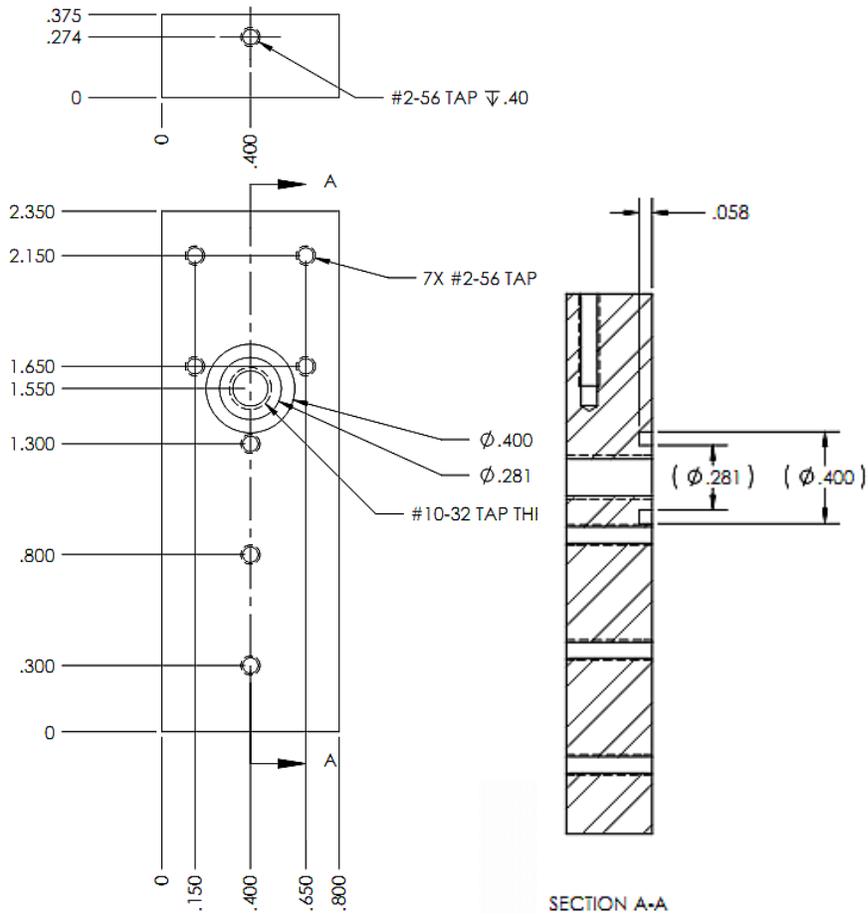

SECTION A-A

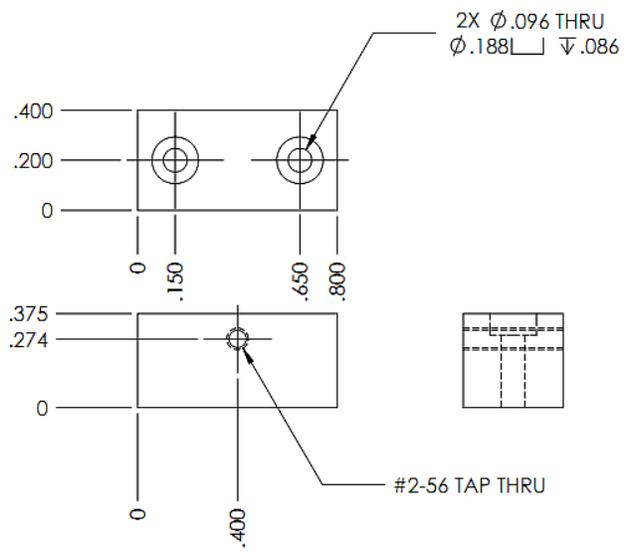

# Experimental Data

| | |
|---|---|
| RIP, N2. Lower curve is (+) | 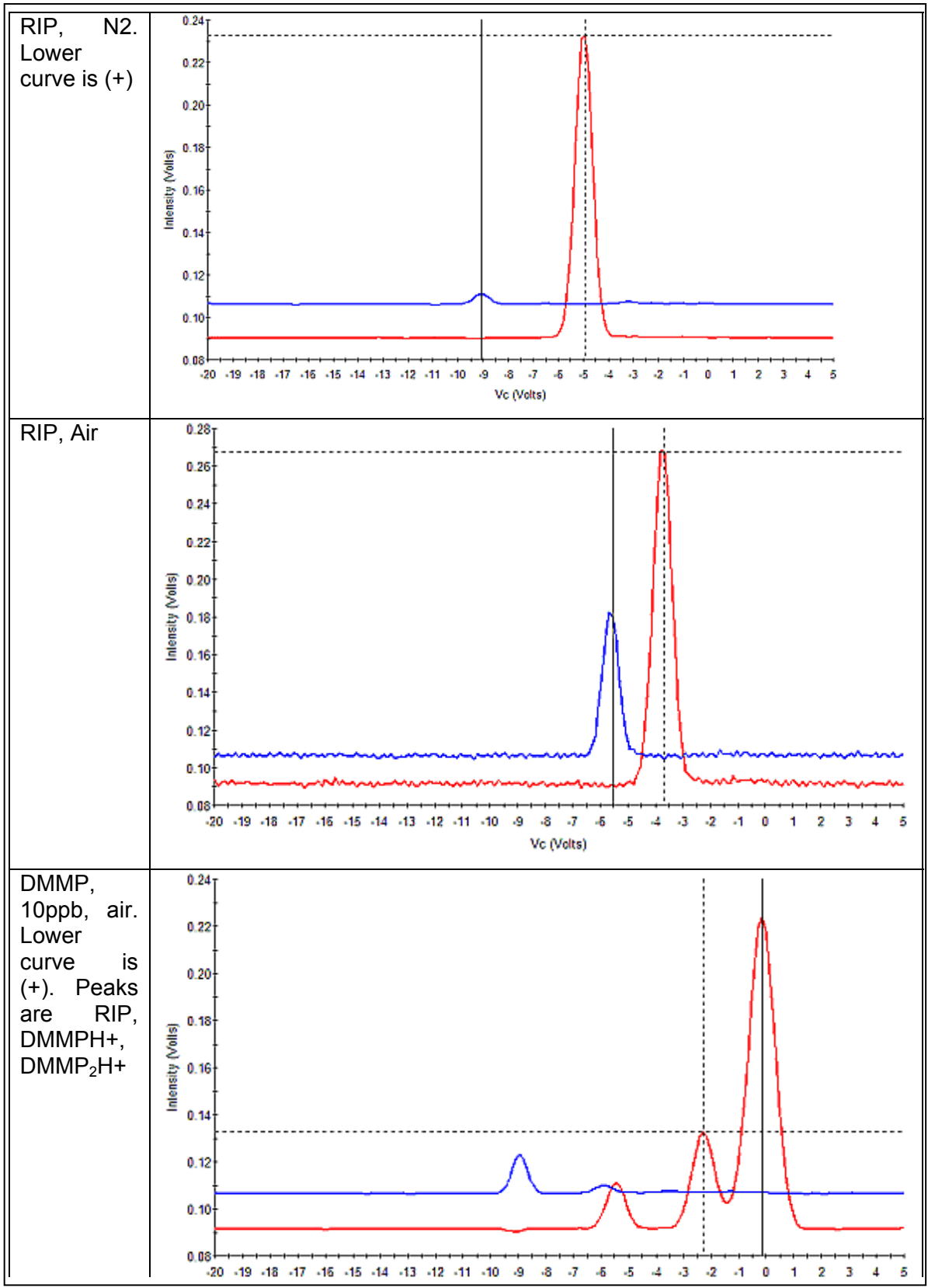 |
| RIP, Air | |
| DMMP, 10ppb, air. Lower curve is (+). Peaks are RIP, DMMPH+, DMMP$_2$H+ | |

**Mass spectra comparing Ni-63 and Crossed-Wires (CGD) sources**

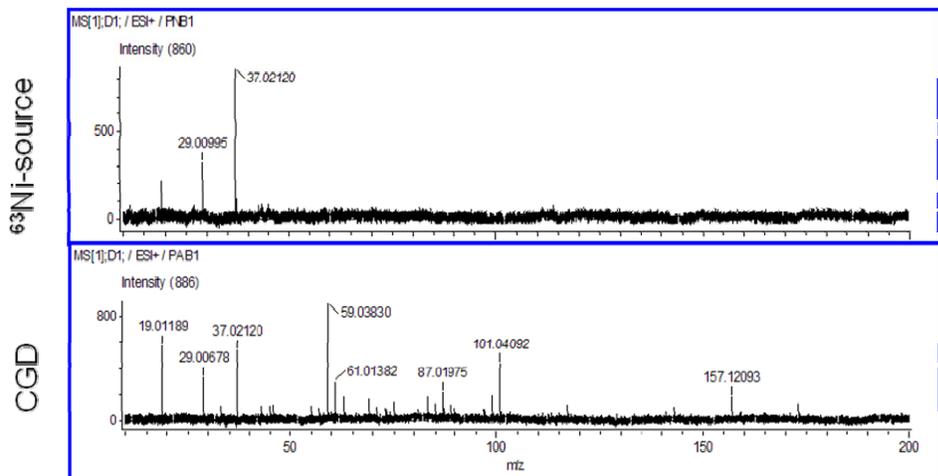
RIP, Air

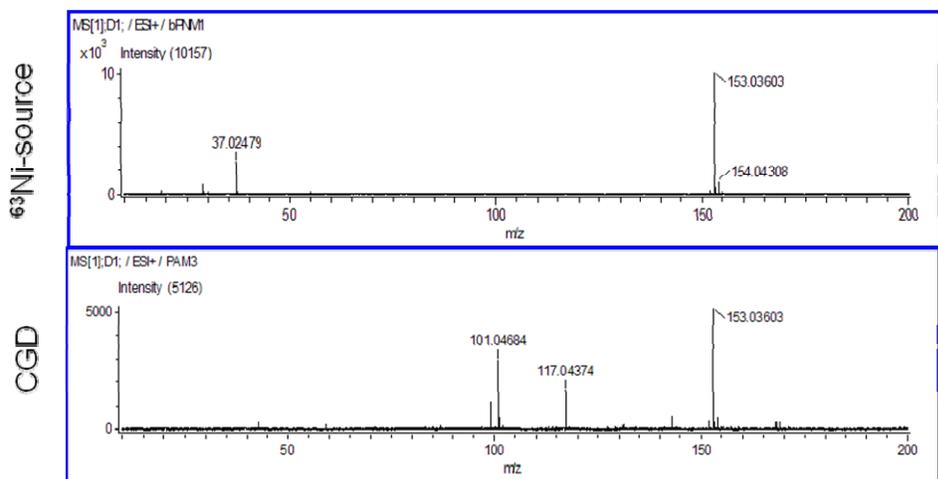
Methyl Salicylate in Air

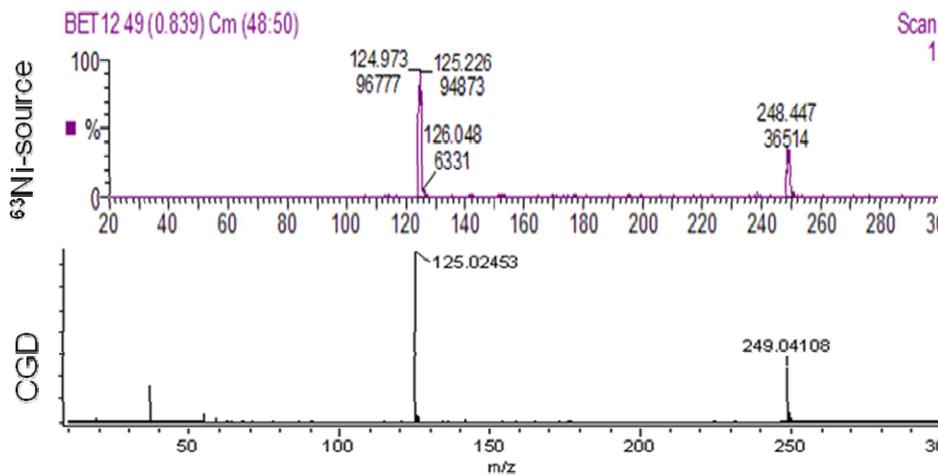
DMMP in Air